\documentstyle[a4,10pt]{article}

\def\beql[#1]         {\begin{equation} \label{#1} }
\def\beq              {\begin{equation} } 
\def\eeq              {\end{equation}   }
\def\beqal[#1]        {\begin{eqnarray} \label{#1} }
\def\beqa             {\begin{eqnarray} }
\def\eeqa             {\end{eqnarray}   }
\def\eeqann           {\nnum \end{eqnarray}}
\def\nnum             {\nonumber}

\def\scap             {\noindent \ul}

\def\ul               {\underline }     % tex abbreviation
\def\r[#1]            { \quad {\rm #1} \quad }

\def\dofs    {degrees of freedom}

\def\lc      {light-cone}
\def\sm      {Schwin\-ger mo\-del}
\def\pe      {periodic}
\def\ap      {antiperiodic}
\def\bc      {boundary conditions}

\def\kd[#1]           {\delta_{#1}}     % Kronecker delta
\def\d,#1             {\partial_{#1}}   % partial derivative

\def\1,#1             {{1 \over {#1}}}
\def\ma[#1,#2,#3,#4]  {{\left( \matrix{ #1  & #2 \cr
                                        #3  & #4 \cr } \right)}}
\def\ve[#1,#2]        {\left( { #1 \atop #2 } \right) }
\def\bra[#1]          {{\big\langle} {#1} {\big|}}
\def\ket[#1]          {{\big|} {#1} {\big\rangle}}
\def\bk[#1,#2]        {{\big\langle} {#1} {\big|} {#2} {\big\rangle}}
\def\ev[#1]           {{\big\langle} {#1} {\big\rangle}}
\def\com[#1,#2]       {{\Big[} {#1} , {#2} {\Big]}}
\def\acom[#1,#2]      {{\Big\{} {#1} , {#2} {\Big\}}}

\def\ferMa   {m_{\it f}}                        % bare fermion mass
\def\bosMa   {m_{\it B}}                        % Schwinger boson mass
\def\cc      {g}                                % the coupling constant g
\def\lct     {x^+}                              % LC time
\def\lcx     {x^-}                              % LC space
\def\xperp   {x_\perp}                          % LC perp directions
\def\lctim[#1] {{#1}^+}                         % LC time of #1
\def\lcspa[#1] {{#1}^-}                         % LC space of #1
\def\zm[#1]  {\langle {{#1}} \rangle _0}        % zero mode
\def\nm[#1]  {\langle {{#1}} \rangle _n}        % normal mode
\def\massev2          {M^2}
\def\mass2            {M^2}
\def\res              {K}

\def\qb               {\bar{q}}

\def\Cq0              {C_{q \to q}^{(0)}        (1)}
\def\Cqb0             {C_{\qb \to \qb }^{(0)}   (1)}
\def\ellcc            {\lambda}

\def\boxLe            {L}

\def\momentum         {P^+}
\def\energy           {P^-}

\def\change[#1,#2]    {c_{{#1},{#2}}}

\def\conjMom[#1]      {\Pi_{#1}}

\begin{document}

%----------------------------------------------------------------------
%                             start
%----------------------------------------------------------------------
\title{
{\bf The Spectrum of ${\rm QED}_{1+1}$ in the framework
of the DLCQ~method} 
\thanks{
invited talk given at the {\em Hadron Structure '94} conference,
Kosice, Slovakia, 19/9 -- 23/9/1994 }
}
\author{
Stephan Elser \\
Max-Planck-Institut f{\"u}r Kernphysik \\
Postfach 10 39 80 \\
69029 Heidelberg 
}

\maketitle
%------------------------------------------------------------------
%                              section1
%------------------------------------------------------------------

\noindent
This work is based on a 
diploma thesis done in 1993/94 at the
Max-Planck-Institute for Nuclear Physics, Heidelberg,
under supervision of Prof. H.C. Pauli, MPIK Heidelberg.

\section{Motivation}
We concern ourselves with ${\rm QED}_{1+1}$ in
the Discretized Light-Cone Quantization
formalism 
suggested by Pauli and Brodsky \cite{pauli1}
as presented in 
Eller, Pauli and Brodsky \cite{epb}
to investigate the influence of
the fermion field \bc\ on the spectrum
and wavefunctions.

The DLCQ approach is chosen because its
relativistic,
non-perturbative treatment
of gauge theories enables us to aim for
hadronic physics, i.e. a solution of ${\rm QCD}_{3+1}$
obtaining spectra and wavefunctions.
The boundary condition question is motivated by the so-called
``background field'' or ``zero mode problem''
generally existing for photons and electrons.
In the case of photons this problem,
despite some recent investigations e.g. by 
McCartor \cite{mccartor} or
Heinzl, Krusche and Werner
\cite{heinzl}, is still unsolved.
Insights quantizating fermion fields 
in the 'toy model' of ${\rm QED}_{1+1}$ can be generalized to the 
fundamental problem.

\section{Method}
Following Dirac \cite{dirac,diraclectures} 
physical \dofs\ are
quantized
on a equal light-cone time $\lct = (x^0 + x^3)/\sqrt{2}=0$ surface
instead of on the usual $t=0$ surface.
To find the dynamical degrees of freedom an analysis of the constraints
using e.g. 
the Dirac-Bergman algorithm \cite{diraclectures, sundermeyer}
is used.
The time evolution operator 
propagating in a generalized time direction $\lct $
is regarded as the Hamiltonian.
This light-front scheme,
somewhat improperly called \lc\ quantisation
(LCQ), offers advantages for 
field theories, mainly a simpler vacuum structure \cite{weinberg}. 
For details, we refer to \cite{overview}.

Pauli and Brodsky \cite{pauli1} realized the
existence of a compactified formulation equivalent to 
lattice gauge theory,
where
space is compactified using \pe\ \bc\ 
for physical observables 
on a finite intervall
$\left[ -\boxLe ,+\boxLe \right]$ in \lc\ longitudinal dimension 
$\lcx = (x^0-x^3)/ \sqrt{2}$ and
$\left[ -\boxLe _\perp,+\boxLe _\perp \right]$ in the
transverse dimensions $\xperp = (x^1,x^2) $
and a
basis of plane waves in momentum space.
The problem can be formulated as a Schr{\"o}dinger 
eigenvalue equation 
for the 
time evolution operator 
or equivalently for the 
Lorentz-invariant mass squared.
Writing the theory in
creation and annihilation operators
representing particles with
positive longitudinal momentum as
elements of a denumerable Fock-Hilbert space
results in a finite matrix equation
\beqa
\bra[i] : 2 \momentum \energy - (\vec{P_\perp} )^2: \ket[j] \bra[j] \ket[\Psi]
= m^2 \bra[j] \ket[\Psi] ,
\eeqann
which can be diagonalized numerically.
Approximations like restrictions of
particle number, in the literature mostly called
{\it Tamm-Dancoff truncation} \cite{tamm-dancoff},
or the construction of effective
potentials are solidly based \cite{paulikyff}.
Real physics of the system can be
investigated at each step 
as the continuum limit of
arbitrarily large volume both analytically or numerically.

Not touching on the still existing
problems of DLCQ (``left movers'', 
``zero modes'',
``renormalization of a 
Hamiltonian \lc\ theory''),
we concentrate on the question still under discussion
whether
the choice of \bc\ of the fields
can have impact on the results of the
theory.
For details of the calculation we refer to 
the thesis of Elser \cite{elsthesis}

We therefore consider three cases:
\begin{itemize}
\vspace*{-1truemm}
\item
anti-periodic boundary conditions for fermions (i.e. no zero Fourier mode)
\vspace*{-1truemm}
\item
periodic boundary conditions for fermions (zero Fourier mode neglected)
\vspace*{-1truemm}
\item
periodic boundary conditions for fermions (zero Fourier mode included)
\end{itemize}
To depict our results, two coupling constants are chosen
for convenience by
\beqa
\lambda^2 = {1 \over 1 + {\pi m_f^2 \over g^2}} 
\r[and] \rho = {m_f \over g},
\eeqann
using the existence of an arbitrary mass scale.
The ultra-violett cut-off 
connected to the periodicity length is
the so-called harmonic resolution $\res $.

\section{Schwinger model}
QED in one space and one time
dimension with massless fermions, i.e. $\ferMa =0$,
the Schwinger model \cite{schwinger},
is known 
from an analytic solution
to allow only uncharged states. This is sometimes
called {\it ``quark'' trapping} \cite{schwingerlectures}. 
The calculated spectrum is that of free bosons,
the ground state with mass $\bosMa $ corresponding
to a quark-antiquark pair, i.e. a meson.
Above that we find a continuum of
two, three, $\dots$ meson states.

${\rm QED}_{1+1}$ proper, also called the massive \sm,
shows a number of stable particles calculated by
Coleman
which at vanishing coupling diverges as exspected.
In the strong coupling limit
the result is three stable particles
for the case considered
\cite{coleman}. 

Basic to the expansion around the massless case
is the proof that this limit is allowed
and leads to the soluble massless \sm\ \cite{kogutsusskind}.
Although this still being under discussion
\cite{paranjape},
the mass squared of the ground state was calculated 
pertubatively to
be linear with the bare fermion mass
\cite{kogutsusskind, kogutlattice}.
Low-lying state masses and 
this behaviour were numerically confirmed by lattice gauge theory 
\cite{kogutsusskind,kogutlattice,kogutlattice2,crewther}.

\section{Results}

\begin{figure}[p]
\begin{minipage}{15cm}
\begin{picture}(15,8.8)
\put(-3.5,-16.5)
{
\includegraphics{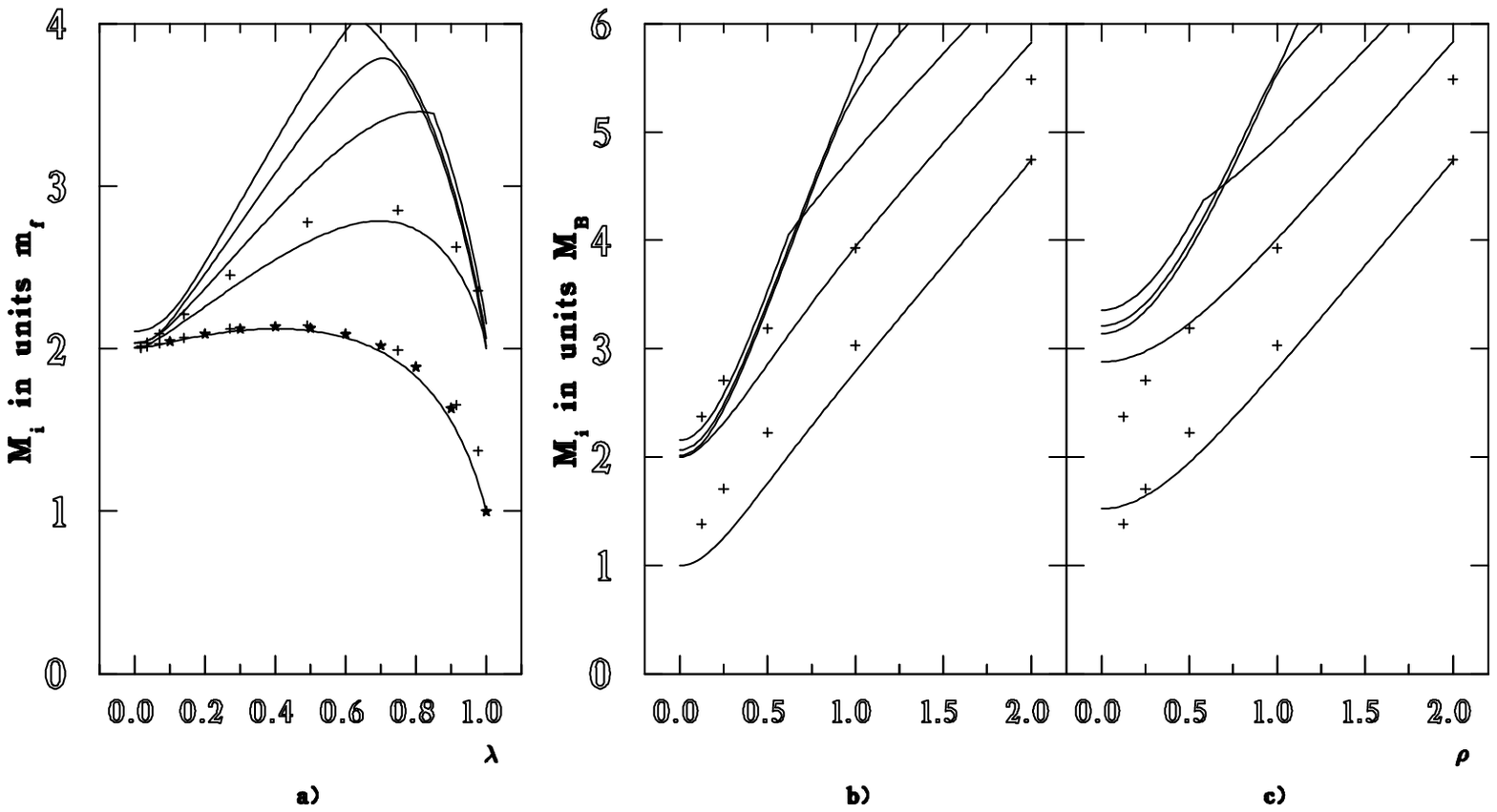}
}
\end{picture}
\end{minipage}
\caption[Comparison to lattice results]
{{\bf Comparison to lattice results.}
\sl 
In 
{\bf a)} results of a DLCQ calculation with \ap\ \bc\ (full lines) 
are compared to lattice gauge (crosses) and 
DLCQ with extrapolation scheme (stars) results. 
The parameter $\ellcc $ is used to depict all mass and coupling
possibilities. \\
In 
{\bf b)} an DLCQ calculation with \ap\ \bc\
is compared to one with 
\pe\ \bc\ neglecting fermion zero modes
in {\bf c)} for low fermion masses corr. to small $\rho$.
}
\label{pic1}
\end{figure}

All results are shown at a fixed
harmonic resolution of $\res =16$ unless extrapolated 
data is mentioned.
Here a quadratic fit routine was used to go beyond this
limit.

\begin{figure}[p]
\begin{minipage}{15cm}
\begin{picture}(15,8.8)
\put(-2.5,-16.6)
{\includegraphics{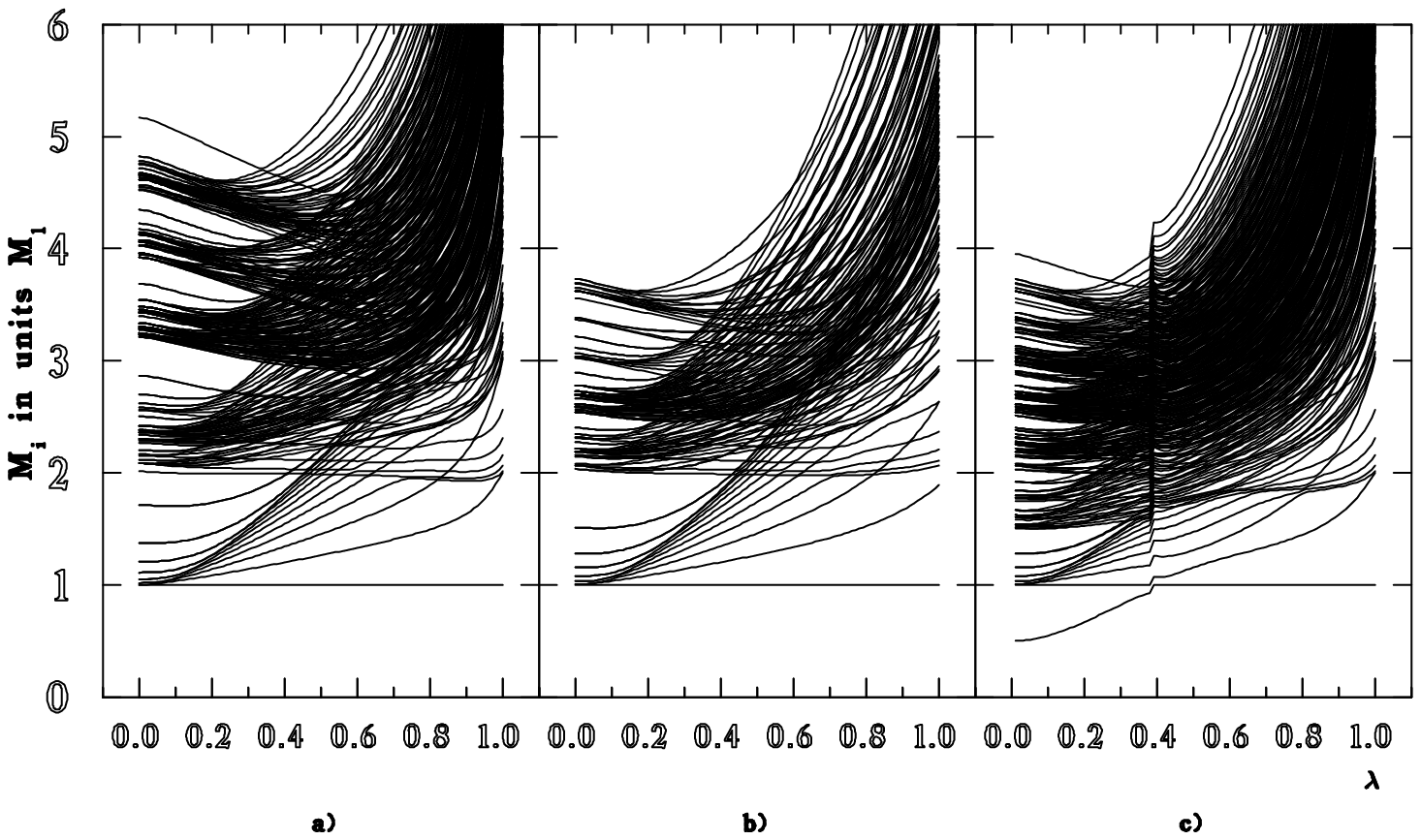}}
\end{picture}
\end{minipage}
\caption[Complete mass spectrum]
{{\bf Complete mass spectrum.}
\sl 
All mass eigenvalues normalized on the physical ground state 
for 
{\bf a)} \ap,
{\bf b)} \pe\ \bc\ neglecting zero modes and
{\bf c)} \pe\ \bc\ incorparating a fermion zero mode 
at a resolution of $\res =16$ against the 
parameter $\ellcc $.}
\label{pic2}
\end{figure}

\scap{Comparison to lattice gauge theory.}
The main impression in fig.\,(\ref{pic1}a) 
considering the lowest states in the spectra is 
the good agreement for the whole range of possible
couplings and masses to the lattice results 
of Crewther and Hamer \cite{crewther}.
Especially the extrapolated data (stars)
is excellent.

In figs.\,(\ref{pic1}b+c)
this is investigated in more detail.
Focussing on the small mass limit,
only the antiperiodic case of fig.\,(2a)
is able to regain the Schwinger ($\ferMa =0$) limit,
whereas the periodic case neglecting zero modes is 50\% off,
showing the importance of the zero mode. 
It also can be
seen that both do not obtain the linear behaviour found in lattice gauge
calculations.

\begin{figure}[t]
\begin{minipage}{15cm}
\begin{picture}(15,6.1)
\put(1.5,-4.35)
{\includegraphics{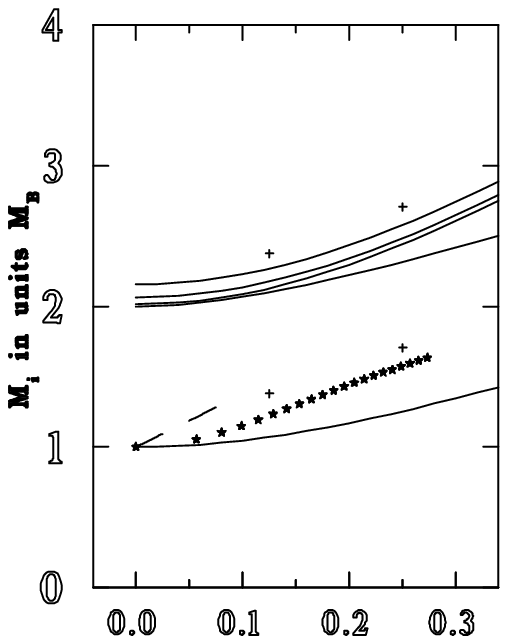}}
\end{picture}
\end{minipage}
\caption[Small mass limit]
{{\bf Small mass limit. }
\sl 
Results of a DLCQ calculation with \ap\ \bc\ (full lines) 
are compared to lattice gauge results (crosses) and 
DLCQ with extrapolation scheme (stars). 
The parameter $\rho $ is used to focus on extremely small 
fermion masses
}
\label{pic3}
\end{figure}

\scap{Complete mass spectrum.}
The mass eigenvalues in fig.\,(\ref{pic2}) are
normalized to
the ground state mass.
Decomposition of scattering states 
with invariant masses below two ground state masses
and bound stable states is then obvious.
States just below threshhold are identified as 
weakly bound dipositronium molecules \cite{epb,elsthesis}.
A qualitative consideration
shows good agreement to Coleman's prediction
of stable states in the cases with no neglected zero
modes in figs.\,(\ref{pic2}a+c).

\scap{Tamm-Dancoff truncation.} 
For space reasons we do not depict results, but simply state that 
a good 
quality of this approximation
in the extreme case of a two-particle cut-off
can be found for small coupling constants,
where all stable states can be thus obtained.
But already at a coupling of $\cc = 0.9\,\ferMa $ ($\ellcc = 0.45$) 
this is no longer true.
Exactly the interesting weakly bound states
are four particle states and
therefore unobtainable in this extreme approximation.

\scap{Small mass limit.}
Recent investigation
based on work by K. Hornbostel \cite{hornbostel} 
showed that the problems in the small mass limit
can be regarded as numerical artefacts from the limited 
harmonic resolution. A extrapolation scheme
is able to improve results by a factor of about 20
using the same computer facilities. This is depicted
is fig.\,(\ref{pic3}) suggesting
that the linear behaviour of the mass squared can be obtained
to a very high precision. 

\section{Extensions}
In this work we did for space reasons 
not include some rather preliminary
investigations into 
ground and exited states wave functions
and structure functions calculated in the full Fock space.
This is different to other approaches, where
only the two-particle sector was investigated,
and results in detailed information on the internal 
structure of bound many-particle states \cite{epb}.
Following a suggestion by L.C.L. Hollenberg \cite{hol94} 
we also calculated finite temperature quantities
like the energy density and the specific heat from
our mass spectra results assuming 
the partition function of a micro-canonical ensemble
\cite{els94b}.

\section{Conclusions}
Mainly, the method of DLCQ and the results of Eller, Pauli and Brodsky 
are confirmed.
We obtained relativistic, non-pertubative mass spectra 
for a gauge field theory with dynamic fermions
and (not shown) wave functions, structure functions and 
thermal quantities.
Tamm-Dancoff particle number cut-offs are possible if treated
with some care.

The basic lesson is that
one Fourier mode (namely the zero mode) can be decisive 
for even the ground state mass.
The choice of boundary conditions is seen to be not important,
if and only if
all degrees of freedom are treated properly.
This is still a problem for light-cone quantization in general.

%----------------------------------------------------------------------
%                             bibliography
%----------------------------------------------------------------------

\end{document}